\documentclass[twocolumn,showpacs,preprintnumbers,amsmath,amssymb]{revtex4}
\usepackage{graphicx}
\usepackage{dcolumn}
\usepackage{bm}
\usepackage{tabularx}
\input{epsf.tex}

\begin{document}

\title{Charge splitting in ${\pi}N\Delta$ formfactor.}
\author{A.~B.~Gridnev,~ V.~V.~Abaev } 

\vspace{0.5 cm}

\affiliation{Petersburg Nuclear Physics Institute,National Reseach Centre
Kurchatov Institute, 188300 Gatchina, Russia}
\date{\today}

\begin{abstract}
Experimental data on $\pi N $ scattering in the elastic energy region  $W \leq  1.45 GeV $ are analyzed 
within the  $K$-matrix approach with effective Lagrangians. The  charge splitting in $\delta^{++}_{33}
$ and $\delta^{0}_{33}$ phases obtaned in the phase shift analisys is studed . It is shown that the change
in the sign in the energy dependence of the phase difference  can be obtained with different 
formfactors of the $\Delta^{++}$ and $\Delta^{0}$ only . This means that $\Delta^{++}$ and $\Delta^{0}$
 has a different size in $\pi N $ interaction.The underlying nature of the observed phenomena is discussed. 

\pacs{14.20.Gk,13.60.Rj,13.60.Le}
\end{abstract}
\thanks{Electronic address: gridnev\_ab@pnpi.nrcki.ru}
\maketitle

Pion-nucleon scattering is one of the fundamental processes that tests low energy quantum 
chromodynamics (QCD)~- the pion is the Goldstone bosone in the chiral limit, where the pion-nucleon interaction
goes to zero at the zero energy. This behavior is modified by an explicit chiral symmetry breaking
by quark masses of the $ m_u \sim 3 MeV$ and $m_d \sim  5 MeV$ ~\cite{pdg}. Since the quark masses are not
equal, Lagrangian of QCD contains a term that breaks the isospin symmetry, which is proportional to the
difference of the quark masses. Therefore, the experimental data on the of isospin symmetry violation
contain important information on the determination of quark masses. Calculations using chiral
Lagrangian~\cite{wei}  and within the framework of the chiral perturbation theory ~\cite{mei} predict
effects of isotopic invariance violation  of the order of 1\% for the pion-nucleon scattering at low
energies. For this reason, special experimental conditions are required to observe the effects of 
isospin breaking, where they are amplified  by kinematics or for other reasons. One example
is the splitting of the mass of the $\Delta(1232)$ resonance ~\cite{aba,gri}. In this case, near the resonance, 
the phase shifts vary very rapidly with the energy, therefore, a small ($\sim 1\%$) difference in 
the masses of the different isotopic components $\Delta(1232)$ of the resonance leads to a significant
difference in the phase shifts.
In paper ~\cite{aba}, the results of the discrete phase analysis of $ \pi N $ elastic scattering data in the 
energy region  $W \leq  1.52 GeV$ were presented. One of the goals of this analysis was a search for the
isospin violation in $ \pi N $ interaction. At the first step, the $ \pi^+ P $ data were analysed only.
Then $ \pi^- P $ and charge exchange data were analysed within the isospin symmetry relation with a soft constrain
from $ \pi^+ P $ analysis. As a result, all phase shifts except $ P_{3/2} $ obtaned in the first step and second 
are equal within the errors.The $ P_{3/2} $ phase shifts from $ \pi^+ P $ analysis ($\delta^{++}_{33}$) and
 $ P_{3/2} $ phase shifts from $ \pi^- P $ and charge exchange analysis ($\delta^{0}_{33}$) are found different
 far from the errors. Then $\delta^{++}_{33}$ and $\delta^{0}_{33}$ were fitted by Breit-Wigner formulae to determine
 the masses of the $\Delta^{++}$ and $\Delta^{0}$ resonances ~\cite{aba,pdg}. The same results were obtained in ~\cite{gri}
 from the analysis of $ \pi N $ scattering data in the resonance energy region $ T_{\pi} \leq 250 MeV $  within 
 a K-matrix approach with effective lagrangians ~\cite{gri,pdg}. However, the phase shift
 analysis~\cite{aba,aba1} shows the new qualitive phenomenon - $\Delta\delta_{33} = (\delta^{++}_{33} - \delta^{0}_{33}$) , as the function of 
 the energy, changes the sign at $w \sim  1.35 GeV$ . This phenomenon cannot be explained with two simple Breit-Wigner formulae 
 with two parameters mass and width.

\begin{figure}[!ht]
  \centering
    \includegraphics[width=0.50\textwidth ]{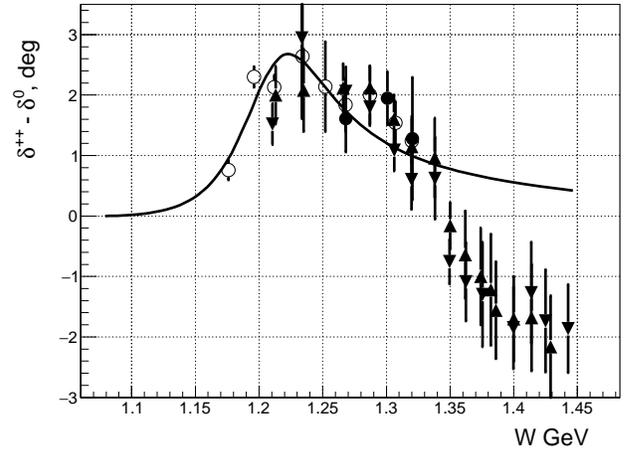}
  \caption{The $\delta^{++}_{33} - \delta^{0}_{33}$  as function of energy .Solid line correspond to fit with two
  Breit-Wigner formulae with parameters from ~\cite{pdg} . Data : triangles up and triangles down are taken from~\cite{aba}
  and ~\cite{aba1}, empty circles from~\cite{bugg} and filled circles from~\cite{alder}. }
  \label{fig:df0}
  
\end{figure}

The goal of this work is to study how this phenomenon can be explaned in K-matrix approach with effective lagrangians. The detailed
description of this approach can be found in~\cite{gri,gri1}. It is assumed that K-matrix, being a solution of
 the Bethe-Salpeter eqation, can be considered as a sum of the tree-level Feynman diagrams with the effective Lagrangians in the
 vertices. The t channel contains an effective $\sigma$ and $\rho$ meson exchahge . In s and u channels we include nucleon and $\Delta$
 isobar exchange. The contribution from the elastic tails of other resonanses are accounted for also. The parameters of these resonances
 are taken from ~\cite{pdg}. As in ~\cite{gri}, the isospin symmetry is not assumed to be explicit and all the calculations were performed in
 the charged channels. The masses of the incoming and outgoing particles were taken as masses of the physical particles from
 ~\cite{pdg}. The masses of the intermediate particles and the coupling constants in the vertices are free parameters and can be 
 different for different charged channels. The free parameters were determined by fiting the results of the calculations to 
 the experimental data.

 In ~\cite{gri}, it was demonstrated that such an approach works very well in the $ \Delta$ resonance energy region
 $ T_{\pi} \leq 250 MeV(W< 1.28 GeV) $. Here we consider the whole elasic energy region $ W< 1.45 GeV $.
 The database contains all the experimental data on elastic $\pi^{\pm}P$ scattering cross sections and charge exchange (CEX) 
up to $ W< 1.45 GeV $ which can be found in  SAID database ~\cite{said} (about 5000 datapoints). A rather good fit was
 obtained with $\chi^2 \sim $ to 1.8 per point. The masses of the $\Delta^{++}$ and $\Delta^{0} $ which we get from this fit
  coincide with ~\cite{gri} within the errors. No statistically proved differences in coupling constants for different
 charged channels were found. For example, $g_{\pi N \Delta^{++}}$ and $g_{\pi N \Delta^{0}}$ are coincides with the accuracy 
 0.2\% . The ($\delta^{++}_{33} - \delta^{0}_{33}$) remained positive for all the enegies. From this we conclude that the fit
  in the whole elastic energy region gives the same results as in ~\cite{gri} around the resonance.
   Let us look how the $\delta^{++}_{33} - \delta^{0}_{33}$  comes out in our approach. The s channel $\Delta$ exchange graph gives
   the main contribution to it. To calculate it we need expressions for vertices and Green function.  The interaction Lagrangian 
   corresponding  to $ \pi N \Delta$ vertex, reads as
   \begin{equation}
   \label{1}
   L_{\pi n \Delta}=\frac{g_{\pi n \Delta}}{2m}\bar \Delta_\mu[g_{\mu\nu}-z\gamma_\mu\gamma_\nu]\ T \partial_\nu  \pi\Psi
   \end{equation}
   
Here, T denots the transition operator between the nucleon and $\Delta$ isobar. The Green function has the form ~\cite{hoh}

   \begin{equation}
   \label{2}
   G_{\mu\nu}=\frac{\hat P+M}{P^2-M^2}(g_{\mu\nu}-\frac{2}{3}\frac{P_\mu 
P_\nu}{M^2})+\frac{1}{3}\frac{P_\mu}{M}\gamma_\nu- 
\frac{1}{3}\frac{P_\nu}{M}\gamma_\mu-\frac{1}{3}\gamma_\mu\gamma_\nu)
   \end{equation}
with the condition

   \begin{equation}
   \label{3}
   P_\mu G_{\mu\nu}=0; ~~~~  for ~~~~P^2=M^2~~only
   \end{equation}

Here, P is the momentum of the $\Delta$ isobar. Due to (3) condition, the second term in (1) equal to zero 
when $\Delta$ isobar is on the mass shell. In the elastic region, K-matrix has the form
   \begin{equation}
   \label{4}
   K(w)=\frac{g^2_{\pi N \Delta}f(w)}{w-M_\Delta}=\tan(\delta)
   \end{equation}

Where $f(w)$ contains the remained part of the direct graph contribution and practically does not depend on $M_\Delta$.
If the coupling constants $g_{\pi N \Delta}$ are equal for $\Delta^{++}$ and $\Delta^{0}$,   we can estimate the
$\delta^{++}_{33} - \delta^{0}_{33}$  as

   \begin{equation}
   \label{5}
   \frac{w-M_{++}}{g^2_{\pi N \Delta}f(w)} -\frac{w-M_{0}}{g^2_{\pi N \Delta}f(w)}=\frac{M_{++}-M_{0}}{g^2_{\pi N \Delta}f(w)}
   \end{equation}
In the resonance region
   \begin{equation}
   \label{6}
\cot(\delta^{++})- \cot(\delta^0)\approx \delta^{++}-\delta^0
   \end{equation}

\begin{figure}[!ht]
  \centering
    \includegraphics[width=0.50\textwidth ]{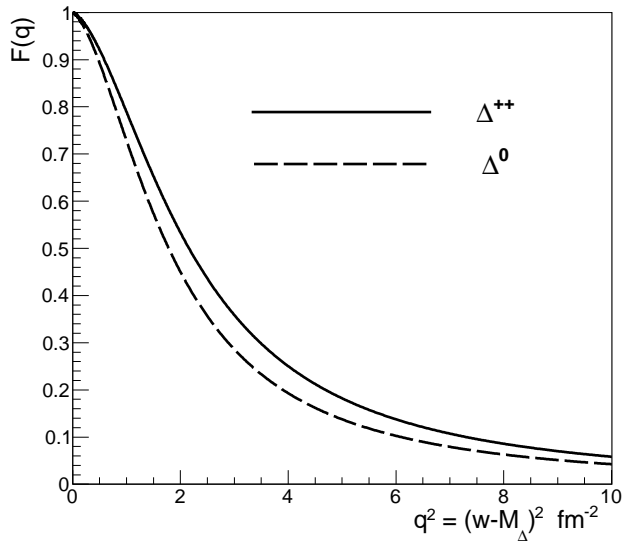}
  \caption{The formfactors F(q)  as the function of $(w-M_\Delta)^2$ for the Green function (10). The solid line corresponds to $\Delta^{++}$
  fomfactor and the dashed line to $\Delta^{0}$.  }
  \label{fig:ff}
  
\end{figure}

Taking into account the relations between the scattering amplitude, K-matrix and Breit-Wigner formula finally gets
   \begin{equation}
   \label{7}
   f=\frac{K}{1.-iK}=\frac{g^2_{\pi N \Delta}f(w)}{1-ig^2_{\pi N \Delta}f(w)}=\frac{\frac{\Gamma}{2}}{1-i\frac{\Gamma}{2}}
   \end{equation}

   \begin{equation}
   \label{8}
   \delta^{++}-\delta^0=\frac{M_{++}-M_{0}}{\frac{\Gamma}{2}}= 3^{\circ}
   \end{equation}
This value is in a good agreement with ~Fig.~\ref{fig:df0}  at the resonance position. From this we conclude that the value of  
$\delta^{++}_{33} - \delta^{0}_{33}$ determinated by the $\Delta$ mass difference. However, from the equation (5) we can see that
$\delta^{++}_{33}$ cannot be equal $\delta^{0}_{33}$ if  $g_{\pi N \Delta^{++}}$ = $g_{\pi N \Delta^{0}}$. Assume
that these coupling constants can be different from

   \begin{equation}
   \label{9}
   \frac{w-M_{++}}{g^2_{\pi N \Delta^{++}}f(w)} -\frac{w-M_{0}}{g^2_{\pi N \Delta^{0}}f(w)}=0
   \end{equation}
we  find that the solution of this equation for $w\approx 1.35 GeV$ can be reached if $g_{\pi N \Delta^{0}}$ is less than
$g_{\pi N \Delta^{++}}$ by $w\approx1.5$ \%. But such a difference is too much to obtain the reasonable fit of the data. 
This result was confirmed by the explisit calculation - fixing all founded parameters an then decreasing 
$g_{\pi N \Delta^{0}}$ by 1.5 \%. The $\chi^2$ goes up by factor 4.
Terefore, in our approach we should modify  the vertexes or the Green function . In ~\cite{andr}, the condition (3) is applied
for the off-shell also for all other resonances. In this approach, a lot of experimental data  on the meson photoproduction can be explained.
For this case, the Green function has the form
\begin{eqnarray}
G_{\mu\nu}=\frac{\hat P+M}{P^2-M^2}(g_{\mu\nu}-\frac{2}{3}\frac{P_\mu 
P_\nu}{P^2})+\nonumber \\ \frac{1}{3}\frac{P_\mu\hat P}{P^2}\gamma_\nu- 
\frac{1}{3}\frac{P_\nu\hat P}{P^2}\gamma_\mu-\frac{1}{3}\gamma_\mu\gamma_\nu)
\end{eqnarray}
This approach saves one free parameter becouse the second term in (1) gives a zero contribution now. Nevertheless, in this case we get
the same results as before (the masses of the $\Delta$ resonances and the coupling constants) with the same $\chi^2$ per point.
The values of $\delta^{++}_{33} - \delta^{0}_{33}$ are very close with the solid line in ~Fig.~\ref{fig:df0}.
As the next step we modify the vertexes by introducing the formfactors with different parameters for $\Delta^{++}$ and $\Delta^{0}$ 
resonances. A relativistic version of the popular "dipole" formfactor was used
\begin{equation}
\label{11}
F(q^2)=\frac{\alpha}{\alpha + (q^2-M_{\Delta}^2)^2}
\end{equation}
The inclusion of the formfactors leads to somewhat decreasing $\chi^2$ per point from 1.8 to 1.7. Clearly different parameters were
obtained for $\Delta^{++}$ and $\Delta^{0}$ formfactors from the fit for both variants of the Green function .
The results are shown in ~Fig.~\ref{fig:ff} and Table 1.

\begin{figure}[!ht]
  \centering
    \includegraphics[width=0.50\textwidth ]{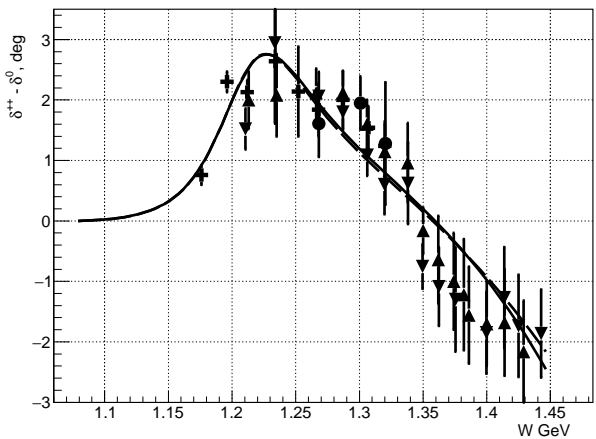}
  \caption{The energy dependence of $\delta^{++}_{33} - \delta^{0}_{33}$. The solid line correspond to 
  the Green function (2) and the dashed line to the Green function (10).}
  \label{fig:dfc}
  
\end{figure}

The calculated values of $\delta^{++}_{33} - \delta^{0}_{33}$ are shown in ~Fig.~\ref{fig:dfc}. From this picture we see that different
 formfactors for $\Delta^{++}$ and $\Delta^{0}$ lead to the excellent fit to the $\delta^{++}_{33} - \delta^{0}_{33}$ data.

  To calculate the radii associated with the founded formfactors  we use a low $ q^2 $ expansion 
\begin{equation}
\label{12}
F(q^2)=1- \frac{1}{6} <r^2> q^2 +...
\end{equation}

The results are presented in  Table 1. As seen from the Table, the mean-square radii are slightly dependent on the Green function form but
 the difference between them is much more stable $ \Delta\sqrt{r^2}= (0.85 \pm 0.14) fm$. The  mean-square radius of $\Delta^{++}$ is found about
  20\% less than that of $\Delta^{0}$ and
both are much  smaller than the charge proton radius $(0.8775 \pm 0.0051)$ fm ~\cite{bern}. It should be noted, however, that radius of the
particle is dependent on the reaction which is used to measure it. In our case, the quark in pion should go inside the proton core to reach the S shell to form the
$\Delta$ isobar. In this sense, our results look  resonable.  

\begin{table}[!ht] \footnotesize
  \centering
	\caption{Formfactor parameters and radii.}
	\label{tab:par}

	\begin{tabular}{|p{1.2cm}|c|c|c|c|}
		\hline
		& \multicolumn{2}{c}{Green function (2)}~&\multicolumn{2}{|c|}{Green function (10)} \\	
			\hline
		& $\Delta^{++}$~~~~	& $\Delta^{0}$	& $\Delta^{++}$	& $\Delta^{0}$\\	\hline
$\alpha ~(Gev^4)$	& 6.08 $\pm$ 0.02 	& 4.06 $\pm$ 0.02 & 8.15 $\pm$ 0.02 	& 5.95$\pm$ 0.02 	\\	\hline
$\sqrt{r^2}~(fm)$             & 0.48$\pm$ 0.01 	& 0.57$\pm$ 0.01 & 0.41$\pm$ 0.01 	& 0.49$\pm$ 0.01\\	\hline
	\end{tabular}
\end{table}

To check how the results depend on the form of the selected formfactor we repeat the calculation with an exponential
formfactor
\begin{equation}
\label{13}
F(q^2)=exp(-(\frac{q^2-M_{\Delta}^2}{\alpha^2})^2)
\end{equation}
As a result~ for the Green function (10)~we get $\sqrt{r^2}$ =(0.46 $\pm$ 0.01) $fm$ for $\Delta^{++}$ and  (0.40 $\pm$ 0.01) $fm $
for $\Delta^{0}$ which is very close to that in Table 1. It is not surprising because our fit was performed in a small region 
$(q^2- M_{\Delta}^2)^2 \approx 0.04 Gev^2$ where the formfactor is practically linear function and does not depend on the  given form.
The main result of this paper - different formfactors (and sizes) of $\Delta^{++}$ and $\Delta^0$ resonances based on the energy dependence
of $\delta^{++}_{33} - \delta^{0}_{33}$ obtained in the discrete phase shift analysis~\cite{aba}. However,   these data have  rather big errors
which are mainly connected with the interpolation experimental data in the resonance region where the data rapidly variate with the energy.
Therefore,
 to decrease the errors, the new experimental data on $\pi^\pm p$ elastic scattering at the same energy are needed as it was done at EPECUR experiment
 ~\cite{epec} at higher energies.

 This work was partially supported by the Russian Fund for
Basic Research grants 15-02-07873.


\end{document}